%% file: Position-Aided_Beam_Prediction.tex
\documentclass[conference,twocolumn]{IEEEtran} 
\input{input.tex}

\usepackage{epstopdf}
\usepackage{enumerate}
\usepackage{algorithmicx}
\usepackage{algorithm}
\usepackage{amsmath}
\usepackage[noend]{algpseudocode}
\usepackage{float}
\usepackage{hyperref}
\usepackage{color}
\usepackage{makeidx}
\usepackage{bbm}
\usepackage{graphicx}
\usepackage{booktabs}
\usepackage{subfigure}

\newcommand{\sref}[1]{{Section}~\ref{#1}}

\newcommand{\argmax}[1]{\underset{#1}{\text{argmax}}}

\begin{document}
\title{Position-aided Beam Prediction in the Real World: \\ How Useful GPS Locations Actually Are?}
\author{João Morais\textsuperscript{1}, Arash Behboodi\textsuperscript{2}, Hamed Pezeshki\textsuperscript{3} and Ahmed Alkhateeb\textsuperscript{1} \\
\textsuperscript{1} Wireless Intelligence Lab, Arizona State University, USA - Emails: \{joao, alkhateeb\}@asu.edu \\
\textsuperscript{2} Qualcomm Technologies Netherlands B.V, The Netherlands, \textsuperscript{3} Qualcomm Research, USA}

\maketitle

\begin{abstract}
Millimeter-wave (mmWave) communication systems rely on narrow beams to achieve sufficient receive signal power. Adjusting these beams is typically associated with large training overhead, which becomes particularly critical for highly-mobile applications. Beam selection can benefit from the knowledge of user positions to reduce the overhead in mmWave beam training. Prior work, however, studied this problem using only synthetic data that does not accurately represent real-world measurements. In this paper, we revisit the position-aided beam prediction problem in light of real-world measurements with commercial-off-the-shelf GPS to derive insights into how much beam training overhead can be saved in practice. We also compare algorithms that perform well in synthetic data but fail to generalize with real data, and attempt to answer what factors cause inference degradation. Further, we propose a machine learning evaluation metric that better captures the end communication system objective. This work aims at closing the gap between reality and simulations in position-aided beam alignment. 
\end{abstract}

\section{Introduction} \label{sec:Intro} 

To unlock the exorbitant data rates available in mmWave frequencies, mobile communications systems are envisioned to use large antenna arrays to combat path loss \cite{itWillWork}. To optimally orient the narrow beams of such arrays, current systems incur in high beam training overhead. High beam training overhead comes from using in-band radio signals to perform channel measurements. This raises the question: Can mmWave communication systems make beam alignment decisions without the expenditure of wireless communication resources? The dependency of mmWave systems on line-of-sight (LOS) links motivated academia to leverage user positions to aid beam alignment \cite{geolocation-side-info,Elizabeth,9425522}. However, to the best of authors' knowledge, position-aided beam alignment is yet to be assessed in real-world scenarios.

Literature holds numerous examples of leveraging user position data to optimize beam alignment. In \cite{geolocation-side-info}, the authors use support vector machines to align beams in a multi-user, multi-cell simulation environment, but consider ideal positions. Likewise, the work in \cite{Elizabeth} uses a deep neural network on synthetic position and orientation data of the user equipment (UE) to reduce the beam alignment overhead. \cite{inverseFinger} proposes a database/lookup table approach for position-aided beam alignment. Although they consider received power noise, the GPS positions are still taken as perfect. The authors in \cite{robertInvited} assess multiple machine learning (ML) methods and consider position errors. However, Gaussian noise may not be suitable for modeling GPS errors in the real-world, since it lacks spatial correlation and can easily be averaged over time \cite{gaussian-noise-is-bad}. 

\vspace{.5cm}

The promising results obtained through simulations raise the question whether similar performance is achievable in the real-world using commercial-off-the-shelf GPS and mmWave communication systems. Yet, real-world studies of position-aided beam prediction are scarce. Knowing how reality differs from simulation represents a crucial step towards closing the gap between academia and industry. With that in mind, this work attempts to answer the following questions:

\begin{itemize}
    \item How do algorithms that work well in synthetic data perform in real-world data? 
	\item What factors degrade the performance of classical position-aided beam alignment methods in real data?
	\item In real data, how much beam training overhead can we save by predicting beams based on GPS positions?
	\item What is a good metric for evaluating beam prediction machine learning models in communication systems?
\end{itemize}

This paper uses channel measurements at 60 GHz and respective GPS positions to answer these questions. The work is organized as follows. Section \ref{system&problem} describes the considered system model and formulates the beam prediction problem. Section \ref{MLsolution} presents the classical algorithms and ML-based solutions to assess. Section \ref{Dataset} exposes our real-world data collection and how data is pre-processed for this work. Section \ref{Results} provides answers to all questions above, with all simulation code available in \cite{GitLink}. 

\section{System Model and Problem Formulation} \label{system&problem}
In this section, we formally state the system model and assumptions, as well as define the beam prediction problem.

\subsection{System Model}
We consider a system where a basestation (BS) with $N$ antennas communicates with a single-antenna UE using one of the $M$ beamforming vectors $\mathbf{f}_m \in \mathbb C^{N\times 1}$ present in its codebook $\boldsymbol{\mathcal F}=\{\mathbf f_m\}_{m=1}^{M}$. If $x$ is the  complex symbol transmitted by the BS using the beamformer $\mathbf{f}_m$, the received symbol in the downlink is given by

\begin{equation}
    y = \mathbf{h}^T \mathbf{f}_m x + n,
\end{equation}

where $\mathbf{h} \in \mathbb C^{N \times 1}$ is the complex channel vector that holds the amplitude and phase transformations that occur between each BS antenna and the UE antenna, and $n \sim \mathcal N_\mathbb C(0,\sigma^2)$ represents complex normally distributed noise. 

\subsection{Problem Formulation}
Adopting receive power as a performance metric, optimal beam selection occurs when the BS picks the beamformer that results in the highest received power $P= \bbE \left[\left|y\right|^2\right]$. Formally, the ideal beamformer $\mathbf{f}^\star$ can be obtained using

\begin{equation} 
    \mathbf{f}^\star = \underset{\mathbf{f} \hspace{.05cm} \in \hspace{.05cm} \boldsymbol{\mathcal{F}}}{\text{argmax}}  \left| \mathbf{h}^T \mathbf{f} \right|^{2}.
\end{equation}

In this work, instead of relying on the explicit channel knowledge, i.e., the knowledge of $\mathbf{h}$, which is hard to acquire, we target predicting the optimal beam based solely on the real-time position information of the UE. If $\mathbf{g} \in \mathbb R^2$ denotes the two-dimensional position vector, composed of latitude $\text{g}_{lat}$ and longitude $\text{g}_{long}$, our problem consists in approximating $\mathbf{f}^\star$ by the estimation $\hat{\mathbf{f}}$ that maximizes $\mathbb{P}(\hat{f}=f^\star|\mathbf{g})$.

\section{Position-Aided Beam Prediction: \hspace{1cm} Proposed Machine Learning Solutions} \label{MLsolution}

Leveraging position for beam prediction is motivated by the directional nature of the narrow beams at mmWave and the higher dependency on Line-of-sight (LOS) paths. The mapping relation (function) between the position and the best beam could arguably be learned using prior observations of position-beam pairs, for example at the infrastructure. In this section we first introduce our problem in light of ML approaches, then we present and justify the choices for such approaches, carefully detailing how they work.

We approximate $\mathbf{f}^\star$ through a prediction function $f_{\Theta}(\mathbf{g})$ parameterized by a set $\Theta$ that represents the parameters of the model. The parameters $\Theta$ are learned from a dataset $\mathcal{D} = \left\{(\mathbf{g}_k, \mathbf f^{\star}_k) : k = 1,..,K \right\}$ which is composed of $K$ labeled training samples. Each sample consists of the input location $\mathbf{g_k}$ and its ground truth optimal beamforming vector $\mathbf f^{\star}_k \in \boldsymbol{\mathcal{F}}$. Therefore, for a given position $\mathbf{g}$ we have $\hat{\mathbf{f}} = f_{\Theta}(\mathbf{g})$. 

The internal behavior of $f_\Theta$ depends on our learning strategy. We use three approaches, respectively, based on a lookup table (LT), K-nearest neighbors (KNN), and a fully connected neural network (NN). The lookup table and KNN approaches represent attractive choices due to their simple and intuitive nature while implicitly integrating spatial correlation knowledge, i.e., they assume similar LOS positions should have similar optimal beams. Lastly, although neural networks have higher complexity, they are powerful inference systems, which can be useful in learning from position data.

All three algorithms estimate a probability distribution $\mathcal P \in \{p_1, \ldots, p_{M} \}$, where $p_m = \mathbb{P}\left(\mathbf{f}_m = \mathbf{f}^\star\right)$. The beamformer with highest probability is chosen:
\begin{equation}
\hat{\mathbf f} = \mathbf{f}_{\widehat{m}},  \ \ \ \widehat{m} = \argmax{m \in \{1, \dots, M\}} p_m.
\end{equation}

We proceed to describe how each beam prediction approach derives the probability vector $\mathcal{P}$.

\subsection{Lookup Table / Position Database}
This approach is motivated by the intuition that a beam covers at least one contiguous physical area. So, we discretize the input space into small areas or cells and attribute a beam to each one. More specifically, given a position $\mathbf{g}$, with each coordinate normalized between 0 and 1, we apply a mapping function $Q(\mathbf{g})$ to find in which cell of a uniform square 2D grid the position falls into. The quantization/cell-mapping function $Q\left(\mathbf{g}\right)$ that maps positions to 2D cells with area inversely proportional to the number of cells, $N_{lt}$, is given by:
\begin{equation}
    Q\left(\mathbf{g}\right) = \left(q_{row}, q_{col}\right) = \left(\left\lfloor\text{g}_{lat} \sqrt{N_{lt}}\right\rfloor, \left\lfloor\text{g}_{lon} \sqrt{N_{lt}}\right\rfloor\right).
\end{equation}

After knowing the cells each training sample belongs to, we attribute a beam to each cell based on the mode of beam reports. In essence, the best beam that is most commonly reported is taken as the top-1 prediction for input positions in that cell, the second most reported beam as top-2, etc. If a cell does not have any training data, the algorithm predicts randomly, and this is becomes more likely as $N_{lt}$ increases and cells get smaller. See Alg. \ref{alg:lt} for a detailed description.

\begin{algorithm}
\caption{Lookup Table}\label{alg:lt}
\begin{algorithmic}
\Require $\mathbf{g}$, $\mathcal{D} = \{\left(\mathbf{g}_k, \mathbf{f}_k^\star\right)\}$, $N_{lt} \geq 1$
\For{$q \gets 1$ to $N_{lt}$}
    \State $\mathcal{U} \gets \left\{k : q = Q(\mathbf{g}_k) \right\}$ \Comment Find samples in cell $q$
    \State $N_q \gets \# \ \mathcal{U}$ \Comment Count samples in cell $q$
    \If{$N_q$ = 0} \Comment If no samples in cell $q$
        \State $\mathcal{P}_q = \left\{p_m = 1/M\right\}_{m=1}^M$ \Comment Predict randomly
        \State \textbf{continue} \Comment{Continue to next cell}
    \EndIf
    \State $\mathcal{F}_{lt} \gets \left\{\mathbf{f}^\star_{u_t} : u_t \in \mathcal{U}\right\}_{t=1}^{N_q}$ \Comment Get labels in cell $q$
    \For{$m \gets 1 $ to $M$} 
        \State $o_m \gets 0$ 
        \For{$f \in \mathcal{F}_{lt}$} \Comment Count occurrences of $f_m$ in $\mathcal{F}_{lt}$ 
            \If{$f = f_m$}
                \State $o_m \gets o_m + 1$
            \EndIf
        \EndFor
    \EndFor
    \State $\mathcal{P}_q \gets \left\{p_m =  o_m/N\right\}$ \Comment Compute probabilities
\EndFor
\State $q \gets Q(\mathbf{g})$ \Comment Find cell of input position
\State $\mathcal{P} \gets \mathcal{P}_q$ \Comment Get prediction for that cell
\end{algorithmic}
\end{algorithm}

\begin{figure*}[!t]
	\centering
	\subfigure[Scenario 1 [2667 datapoints\text{]}]{\includegraphics[width=0.3235\linewidth]{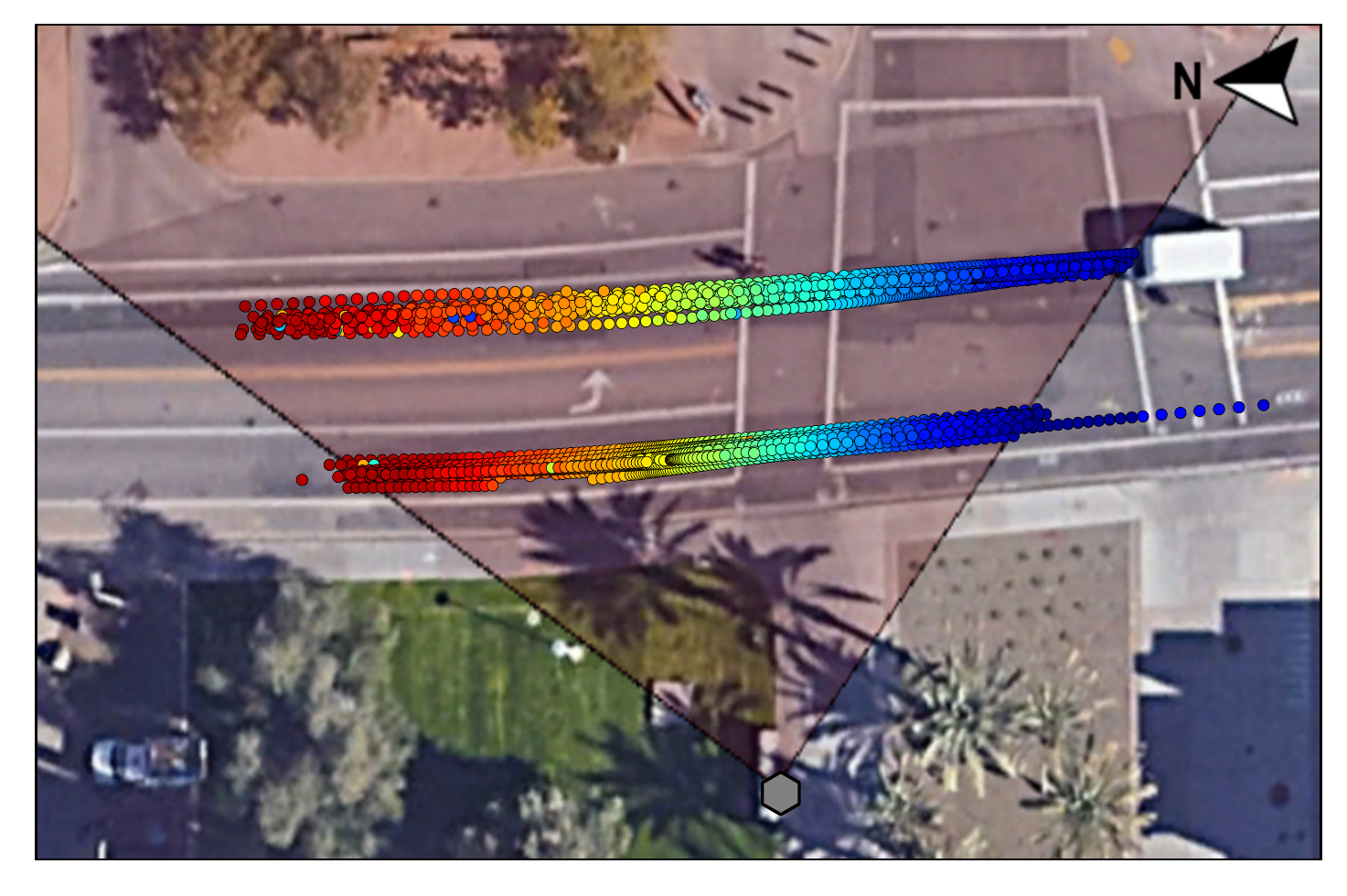}}
	\subfigure[Scenario 6 [1011 datapoints\text{]}]{\includegraphics[width=0.321\linewidth]{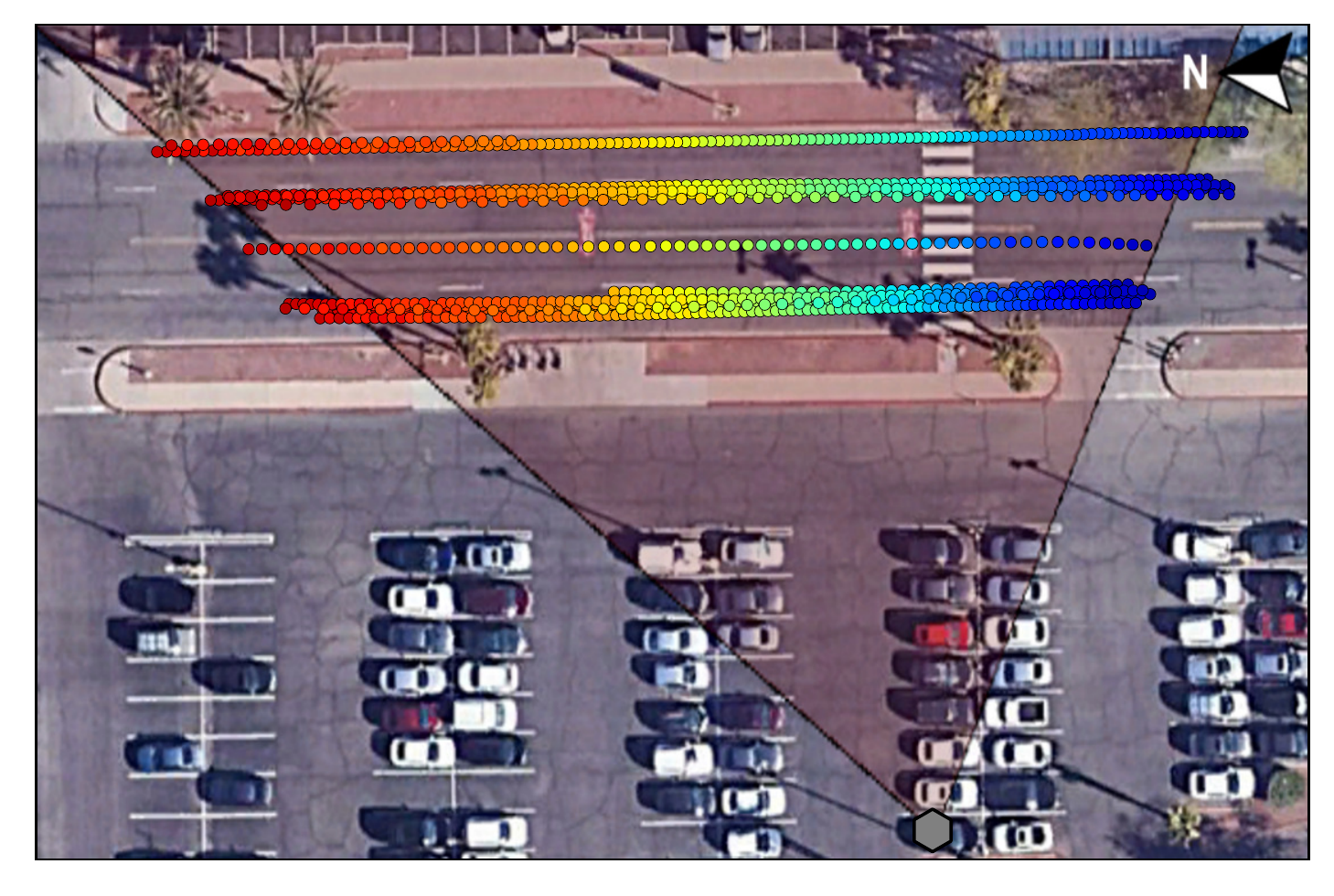}}
	\subfigure[Scenario 7 [897 datapoints\text{]}]{\includegraphics[width=0.327\linewidth]{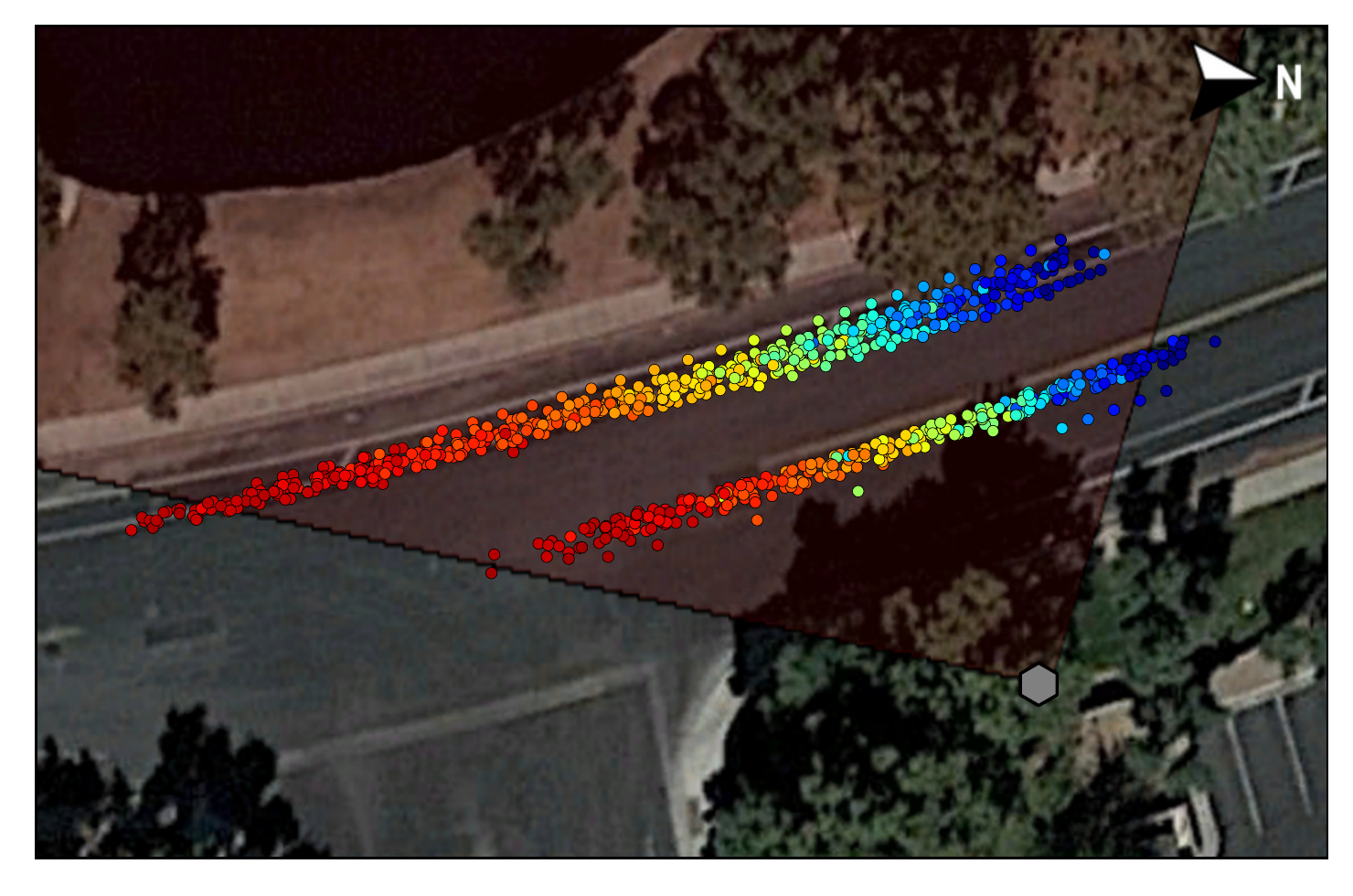}}
	\caption{Color-coded data representation, best beam versus position. The red area represents the Field of View of the RGB camera modality (not used).}
	\label{fig:data}
	\vspace{-3mm}
\end{figure*}

\subsection{K-Nearest Neighbors}
The rationale behind KNN is that similar/neighbor positions should have similar beams. We first identify the $N_{knn}$ nearest neighbors. The selected neighbors then report their best beam. We take the mode of those reports as our best beam estimation. See Alg. \ref{alg:knn} for the detailed steps.
\vspace{-.09cm}
\begin{algorithm}
\caption{K-Nearest Neighbors}\label{alg:knn}
\begin{algorithmic}
\Require $\mathbf{g}$, $\mathcal{D} = \{\left(\mathbf{g}_k, \mathbf{f}_k^\star\right)\}$, $N_{knn} \geq 1$
\State $\left\{d_k \gets \left|\mathbf{g}_k - \mathbf{g}\right|\right\}_{k=1}^K$ \Comment distances to all training samples
\State Let $\left\{u_n\right\}_{n=1}^{N_{knn}}$ be the indices of the $N_{knn}$ smallest distances
\State $\mathcal{F}_{knn} \gets \left\{\mathbf{f}^\star_{u_n}\right\}_{n=1}^{N_{knn}}$ \Comment Get the labels for closest neighbors 
\State Let $\left\{o_m\right\}_{m=1}^M$ be the number of occurrences of $f_m$ in $\mathcal{F}_{knn}$
\State $\mathcal{P} \gets \left\{p_m =  o_m/N\right\}$
\end{algorithmic}
\end{algorithm}

\subsection{Neural Network}

We use a standard fully-connected NN setup. The architecture and main hyperparameters are present in Table \ref{tab:nn}. We found this architecture to be the inflection point after which further increasing the model complexity leads to no returns. Note that one hidden layer with 128 nodes gets us just 2.5\% less accuracy on average. Therefore, we are certain this network size is not limiting our results. 

For NN training, after training for 60 epochs, we see in which epoch the network had the highest accuracy on the validation set and use that model to assess accuracy on the test set. We use cross-entropy loss and tune the neural network weights with Adam optimizer. The initial learning rate provided to Adam is 0.01 and we use a multi-step learning rate scheduler that multiplies all learning parameters by 0.2 at epochs 20 and 40. Finally, as an additional pre-processing step for the neural network, we quantized the input data in 200 bins, i.e., a resolution of 0.005 considering the normalized input has values between zero and one. 

\section{Measurements and Dataset Description} \label{Dataset}
The data used in this work is part of DeepSense \cite{DeepSense}, an extensive multi-modal real-world dataset constructed by the Wireless Intelligence Lab. The dataset  merges co-existing GPS locations, camera images, radar, lidar, and beam-training power/CSI measurements in each data point. For this work, we use the calibrated GPS positions and power data of DeepSense Scenarios 1 to 9. This section presents how the dataset was acquired and what pre-processing steps we performed on it. 

\subsection{Data acquisition}
DeepSense contains data acquired on different locations around the ASU campus, day and night. The UE is equipped with a GPS inside a moving car and a mmWave omni-directional transmitter operating at 60 GHz. The BS is fixed and equipped with a mmWave receiver and a uniform square array with 64 antenna elements. The car passes in front of the BS in different directions. At an approximate rate of ten times a second, the BS sweeps the codebook $\mathcal{F}$ and samples the powers received in each beam. In Figure \ref{fig:data}, we show for some scenarios a GPS image from above, and a scatter plot of positions, where color indicates the best beam for that position, out of 64 possible beams that sweeping azimuth angles. 


\begin{table}[!t]
	\caption{Architecture and Training Hyper-parameters}
	\centering
	\setlength{\tabcolsep}{5pt}
	\renewcommand{\arraystretch}{1.2}
	\begin{tabular}{@{}l|cc@{}}
		\toprule
		\toprule
		\textbf{Parameters}                     & \textbf{Values}  \\ 
		\midrule \midrule
		\textbf{Input size}                     & 2 ($\text{g}_{lat}$ and $\text{g}_{long}$)\\
		\textbf{Hidden Layers}                  & 3 layers, 256 nodes each \\
		\textbf{Output size}                    & $M \in \{8,16,32,64\}$ \\
		\textbf{Activation}                     & ReLU on hidden layers  \\
		\midrule
		\textbf{Training Batch Size}            & 32                  \\
		\textbf{Learning Rate}                  & $1 \times 10 ^{-2}$  \\
		\textbf{Learning Rate Decay}            & epochs 20 and 40    \\
		\textbf{Learning Rate Reduction Factor} & 0.2                 \\
		\textbf{Total Training Epochs}          & 60                  \\ 
		\bottomrule \bottomrule
	\end{tabular}
	\label{tab:nn}
	\vspace{-4mm}
\end{table}

\begin{figure*}[!t]
	\centering
	\subfigure[Top-1 accuracy comparison on every Scenario.]{\includegraphics[width=1.08\columnwidth]{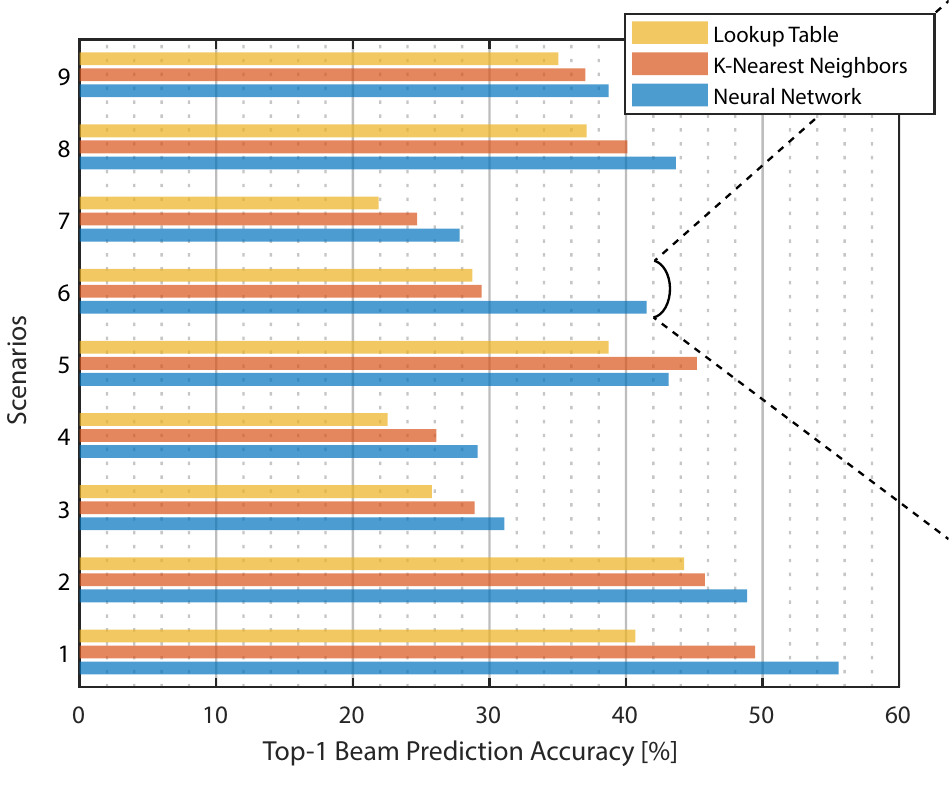}\label{fig:comparisonA}}
	\hspace{-0.015\linewidth}
	\subfigure[Scenario 6 prediction maps.]{\includegraphics[width=0.94\columnwidth]{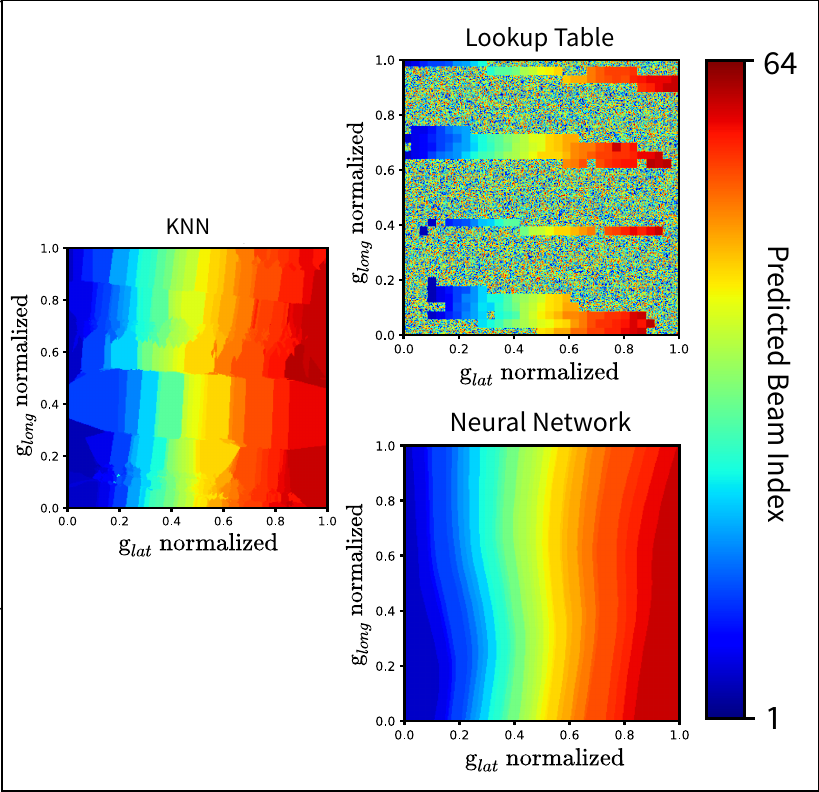}\label{fig:comparisonB}} 
	\label{fig:comparison}
	\caption{Accuracy and beam prediction map comparison for the three considered algorithms, Lookup Table, KNN, and the Neural Network.}
	\vspace{-3mm}
\end{figure*}

\subsection{Data normalization}
Normalization is a standard procedure in machine learning model training. The specific normalization procedure, however, depends on the data and application. Here, we normalize latitudes/longitudes with a min-max normalization on the absolute  coordinates of the UE. We selected this approach after comparing several alternatives, such as using the relative coordinates to the BS, with divide-by-max normalizations that would not distort the scale of the data, and with the usage of polar coordinates, i.e., distance and angle. But the simple min-max normalization always provided the best results.

\subsection{Data split}
To train our models, we split the data of each scenario as follows: 60\% training, 20\% validation, and 20\% testing. It is a common practice to have equal splits for validation and test sets. We tested several similar data splits, and the influence on performance is minimal. 

\section{Real-World Experimental Results} \label{Results}

In this section, we use the aforementioned real-world dataset to test the performance of different algorithms and justify why differences between synthetic and real-world results based on properties of the data. We further suggest meaningful performance metrics for real-world operation, we assess the impact of codebook size in prediction accuracy, and we compute overhead savings for different outage probabilities. 

\subsection{Algorithmic comparison} 
First, we compare the performance of the three adopted ML approaches and answer whether using a NN is excessive or justified. In Figure \ref{fig:comparisonA}, we illustrate the top-1 accuracy for the three algorithms in all scenarios. We see the NN consistently outperforming KNN and the LT. We can draw some insights into why this may happen by looking at Figure \ref{fig:comparisonB}. It shows the output (color representing beam index) for each possible input. To generate it, we take the ML models trained for Scenario 6 and use them to predict the beams of 10 thousand uniformly spaced samples that cover our input space. We see that LT and KNN cannot generalize as well as the NN. This is justified because their input parameters (the number of bins/rows in the table and the number of neighbors in KNN), may not be optimal for all positions in a given scenario, depending heavily on the training data. Nevertheless, for each scenario, we used the optimal setting for that parameter. Therefore, the NN is favored further, because it was up against the optimal versions of LT and KNN, and because it does not involve any scenario-specific optimization. 

\begin{figure}
	\centering
	\includegraphics[width=0.9\columnwidth]{"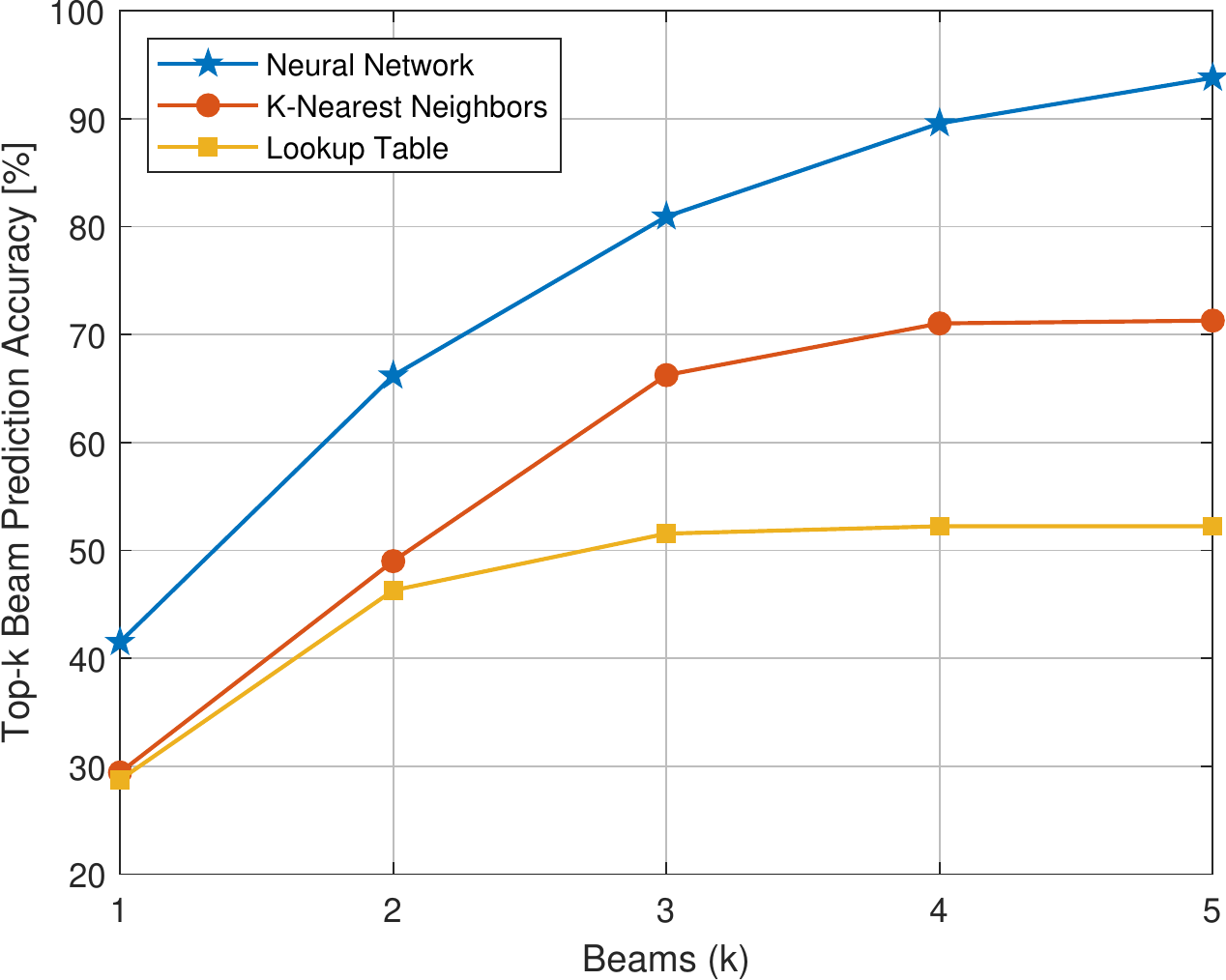"}
	\caption{Top-k accuracy comparison between algorithms for Scenario 6.}
	\label{fig:scen6-knn-vs-nn}
\end{figure}

For systems that rely on ML models to narrow down the set of beams for over-the-air beam training, looking at top-1 prediction is not a complete and infallible way of comparing algorithms. In particular, algorithms may be performing comparatively well in top-1 but fail to generalize in top-3 or top-5 beam prediction accuracies. That is precisely represented in Figure \ref{fig:scen6-knn-vs-nn} where we illustrate how the considered algorithms perform their five most confident guesses of the best beam. We see the NN returning more sensible second-guesses than KNN and LT. KNN and LT approaches fail to generalize to the depth of the NN because they do not consider data as a whole: KNN considers only some close neighbors and LT weighs samples only in a quantized cell. In opposition, the NN is a highly non-linear approach that implicitly uses information for all samples. After all, the network weights are adjusted for every training sample (or batch of samples), even if just infinitesimally. The NN is using and learning from more information than the KNN and LT counterparts, which justifies the quality of its generalizations over KNN and LT.

\subsection{Factors that degrade performance}   \label{subsec:factors}
The results of the previous subsection show significantly different performances in each scenario, and one naturally wonders why. Next, we group the key reasons in two categories, leaving the detailed analysis for follow-up work: 

\begin{itemize}
    \item \textbf{Noisy inputs}: due to GPS inaccuracies or unforeseen latency between the instant of measurement and instant of utilization of said measurement, the reported position can be differ from reality;
    \item \textbf{Noisy labels}: optimal beams may vary over time for the same spatial position, e.g., due to multipath fading, receiver noise, or overlapping of beam patterns. Moving scatterers may further unexpectedly affect propagation, e.g., causing blockages or unforeseen reflections.
\end{itemize}

The aforementioned factors lead to non-intuitive data that is harder to learn. If the position is noiseless and the labels are perfect then we may expect distinctively high prediction accuracy, which is consistent with the results based on synthetic data. We may postulate that low accuracies that we obtain with our real-world dataset were not predicted in literature because authors only considered input noise, which only in part accounts for performance degradation. To overcome this challenge, other auxiliary sources of information, possibly from other sensors, or more complex models and analysis techniques may further enhance the performance.

\begin{figure}
	\centering
	\includegraphics[width=0.9\columnwidth]{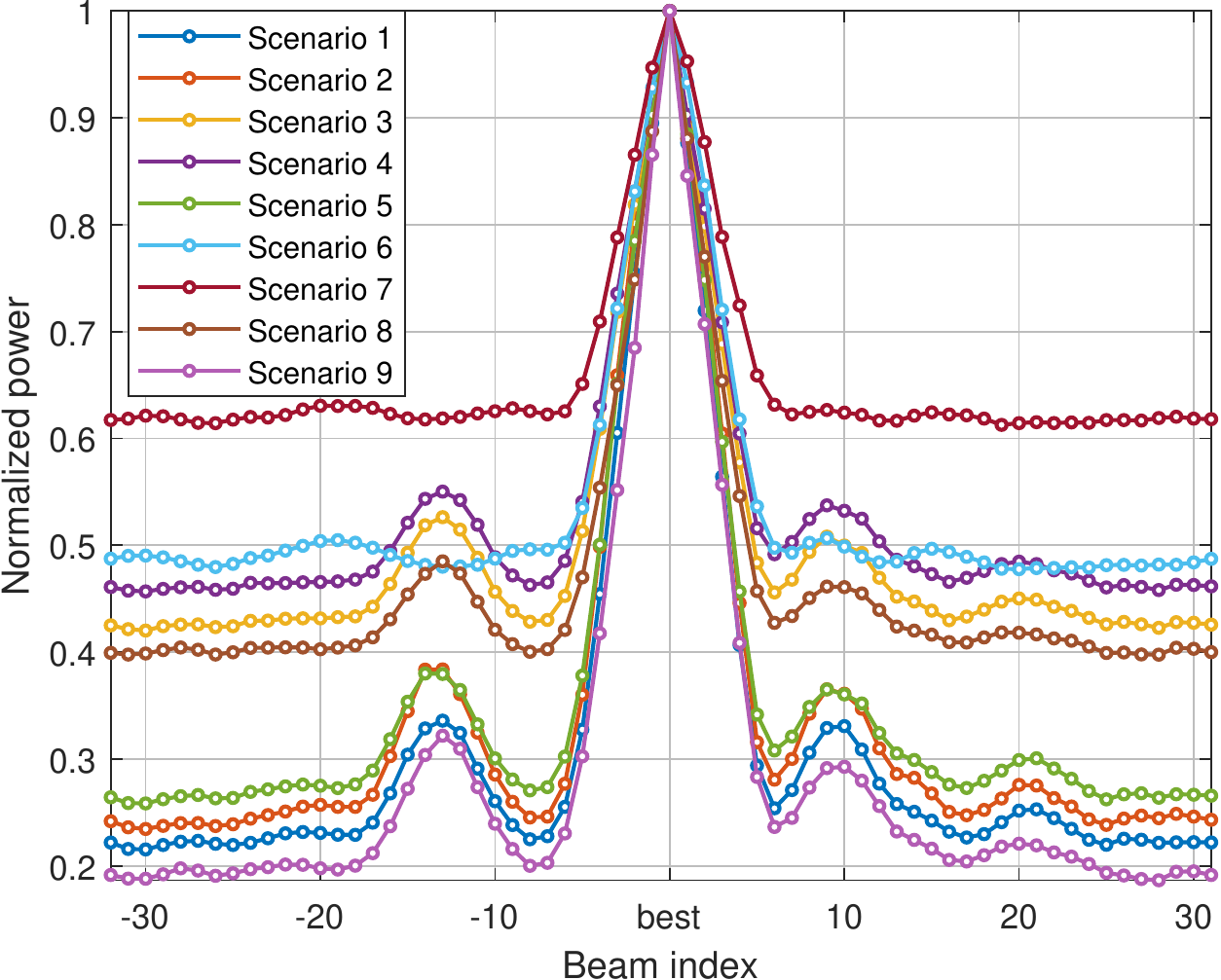}
	\caption{Centered average beam power footprint: the time average of the received power using 64 different beams.}
	\label{fig:pwr_footprint}
	\vspace{-.5cm}
\end{figure}

\subsection{Meaningful performance metrics} 
The  factors discussion in \sref{subsec:factors} affect both system performance and model accuracy, which are often treated as the same quantities. When several beams are received equally well (with similar receive power), selecting any of them results in almost equal performance. However, in terms of model prediction accuracy, not selecting the optimal beam results in clear degradation in this accuracy. To assess how power is distributed across beams, we plot the average power footprint for each scenario in Figure \ref{fig:pwr_footprint}. The power footprint consists of the received powers in all beams, and we average it across all samples taken in each scenario. It tells us how much the accuracy might change across scenarios without a direct impact on system performance, since scenarios with a wider power spread across beams might present lower accuracy, without impacting in performance. We see that the power footprint varies noticeably, suggesting scenarios with not so similar accuracies may have similar performances, hence motivating the employment of more robust metrics.

\begin{table}[!t]
	\caption{Statistics of NN inference in all scenarios. \\Top-1 Accuracy, Power loss, and 70\%-Power Beamset Size.}
	\centering
	\setlength{\tabcolsep}{5pt}
	\renewcommand{\arraystretch}{1.2}
	\begin{tabular}{c|c|c|c} 
		\toprule
		\toprule
		Scenario &    Acc. [\%] & $P_L$ [dB]  & $BS^{0.7}$ \\
		\midrule
		1        & 55.57        & 0.27 & 6.28    \\
		2        & 48.86        & 0.43 & 7.08    \\
		3        & 31.09        & 2.05 & 10.02   \\
		4        & 29.14        & 2.46 & 13.13   \\
		5        & 43.12        & 1.04 & 6.21    \\
		6        & 41.51        & 0.36 & 15.69   \\
		7        & 27.82        & 1.56 & 27.58   \\
		8        & 43.65        & 1.22 & 9.58    \\
		9        & 38.73        & 2.63 & 5.99    \\
		\bottomrule \bottomrule
	\end{tabular}
	\label{gamma_table}
	\vspace{-4mm}
\end{table}

Let us define a metric that targets system performance with more rigor. One such metric is the average power loss between prediction and ground truth, defined as
\begin{equation}
    P_{L[dB]} = 10 \log_{10} \left( \frac{1}{K} \sum_{k=1}^K{\frac{P^k_{\mathbf{f}^\star} - P_{n}}{P^k_{\hat{\mathbf{f}}} - P_{n}}}\right),
\end{equation}
where $P_n$ is the noise power of the scenario, $P^k_{\mathbf{f}^\star}$ is the power of the ground truth beam in sample $k$ and $P^k_{\hat{\mathbf{f}}}$ is the power with our predicted beam for sample $k$. The noise power is the average of the smallest power per sample, i.e, the lowest value of each curve in Figure \ref{fig:pwr_footprint}.

Now, we capture the width of the power footprint plots into a single number we call the Beamset Size (BS), such that $BS^{(0.7)}$ consists of the number of beams within 70\% of the power of the best beam. Our objective is to evaluate how inference accuracy and power relate to $BS^{(0.7)}$. We define $BS^{(0.7)} = \sum_{k=1}^K b_k^{(0.7)}$, with $b_k^{(0.7)} = \sum_{m=1}^M a_{k,m}^{(0.7)}$ and
\begin{equation}
    a_{k,m}^{(0.7)} = 
        \begin{cases}
        1, & P^k_{f_m} \geq 0.7 \ P^k_{f^\star} \\ 
        0, & \ \ \ \text{otherwise} 
        \end{cases},
\end{equation}
where $P^k_{f_m}$ denotes the power received in sample $k$ using the beamformer $f_m$. Now, we show in Table \ref{gamma_table} the relation between the accuracy of each scenario, the 70\%-beamset size, and power loss. First and foremost, note that our solution achieves always less than 3dB power loss. Moreover, note what happens for similar power losses, which represent similar system-level performances. Take, for example, Scenarios 1 and 6, or Scenarios 7 and 8. Their accuracies vary substantially and in accordance with $BS^{(0.7)}$.

More generally, the columns of accuracy and power loss correlate well (-0.7), which makes sense since the system's performance is significantly dependent on the performance of the inference model. However, while accuracy correlates well with $BS^{(0.7)}$ (-0.6), power loss does not relate at all (0.005). We may conclude that $P_L$ is a better system performance metric because it can see beyond the ML task performance since it is not affected by how wide the power footprint is.

\subsection{Accuracy depends on the size of the codebook} 

Although we showed that accuracy does not entirely reflect the overall system performance, it still allows comparison between model, scenarios, and assessing model generalization. The model performance depends directly on how many categories it needs to classify. If the model needs to differentiate between fewer beams, the task is more manageable, and vice-versa. In Table \ref{n_beams_table}, we show quantitatively how model accuracy varies with the number of beams considered. Although values depend to a large degree on the considered scenario, in any scenario it is always easier to classify the best beam among fewer possible options. 

\begin{table}[!t]
	\caption{Top-1 Accuracy comparison with the number of beams M.}
	\centering
	\setlength{\tabcolsep}{5pt}
	\renewcommand{\arraystretch}{1.2}
	\begin{tabular}{c|c|c|c|c} 
		\toprule
		\toprule
        Scenario & M=64  & M=32  & M=16  & M=8   \\
		\midrule
        1        & 55.57 & 71.34 & 86.17 & 90.24 \\
        2        & 48.86 & 60.02 & 78.99 & 88.05 \\
        3        & 31.09 & 37.63 & 55.66 & 70.10 \\
        4        & 29.14 & 35.53 & 52.81 & 69.12 \\
        5        & 43.12 & 55.91 & 74.73 & 84.02 \\
        6        & 41.51 & 63.91 & 79.43 & 90.63 \\
        7        & 27.82 & 41.82 & 62.53 & 76.23 \\
        8        & 43.65 & 56.67 & 66.72 & 76.97 \\
        9        & 38.73 & 49.42 & 66.75 & 76.22 \\
        \midrule
        Average  & 39.9  & 52.5  & 69.3  & 80.2  \\
		\bottomrule \bottomrule
	\end{tabular}
	\label{n_beams_table}
	\vspace{-4mm}
\end{table}

\subsection{How much overhead do GPS positions save?} 

If we define overhead savings as the reduction in the number of beams that require training, then we can guarantee an overhead level that depends only on the reliability (or outage probability) we want in the ML task. The reliability is the confidence we have that a specific set of beams contains the best. Therefore, the higher the confidence, the more beams our set should have, and the lowest our overhead savings. 

\begin{figure}
	\centering
	\includegraphics[width=0.85\columnwidth]{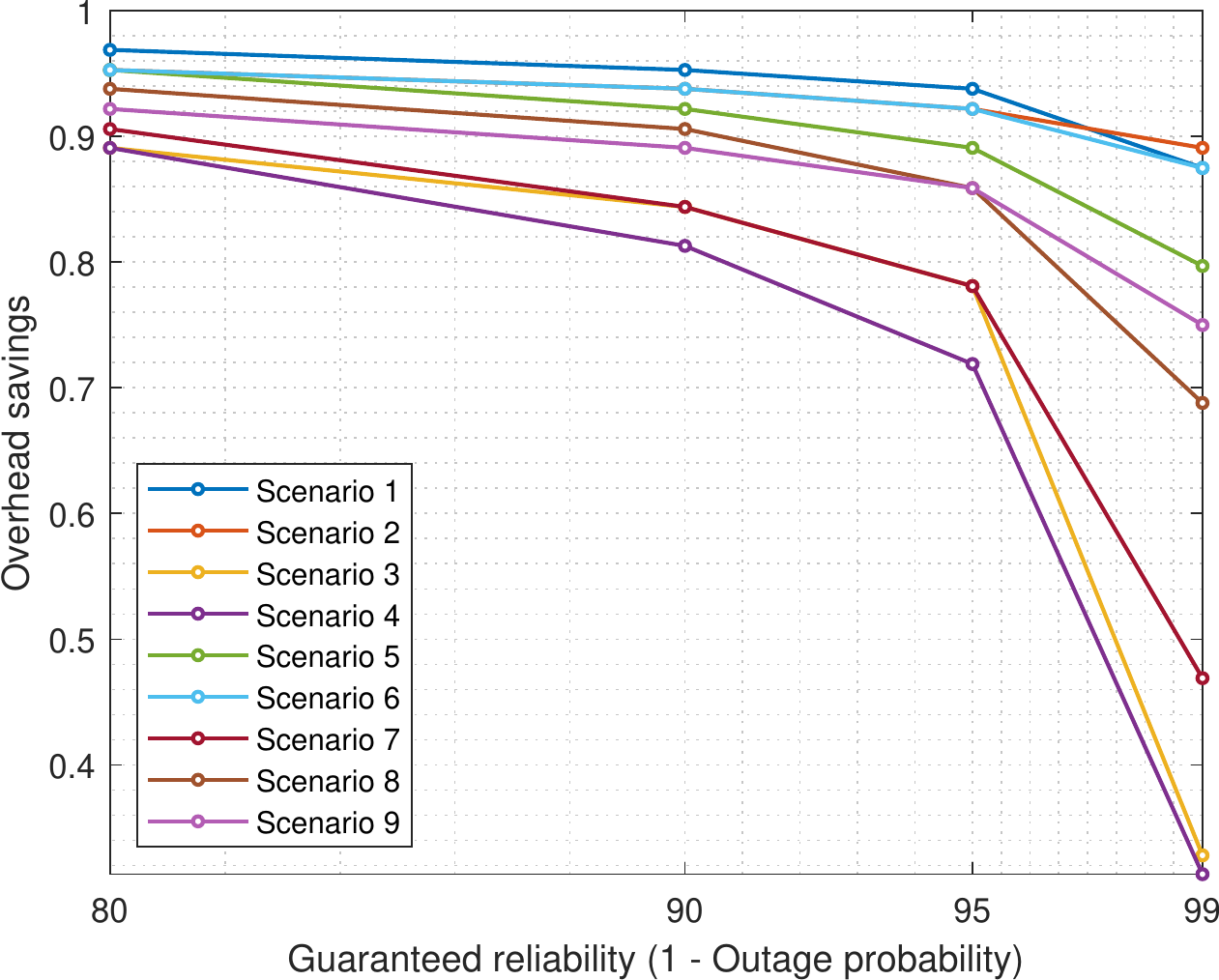}
	\caption{Overhead savings as a function of outage probability.}
	\label{fig:outage}
\end{figure}

Let us assume that all 64 beams would require training for the system to discover the ideal candidate. Then our overhead savings with a reliability of at least 90\%, $\alpha_{OH}^{(90)}$, is given by 

\begin{equation}
    \alpha_{OH}^{(90)} = 1 - \frac{b^{(90)}}{M}
\end{equation}

where $b^{(90)}$ is the minimum number of beams (on average) needed so their likelihood probabilities sum to more than 90\%, and $M$ is the codebook size. For $M=64$, we show in Figure \ref{fig:outage} the overhead savings for typical reliability values. For example, with an reliability of 90\% (outage probability of 10\%), we may save between 81\% and 95\% of training overhead, depending on the scenario. And for 99\% reliability, our approach saves 66\% overhead on average.

\section{Conclusions} \label{Conclusion}

In this work, we conducted real-world experiments with off-the-shelf hardware to assess how much position data could help reducing the beam alignment overhead. We evaluated three algorithms, a lookup table, K-nearest neighbors, and a neural network. Despite intuition favoring the first two, we verified that the neural network was the most capable in dealing with real-world noise, thus justifying its added complexity. But the neural network still showed lower beam prediction accuracies than simulations, so we identified the causes of performance deterioration, and classified them in input noise and label noise. Instead of accuracy, we proceeded to propose a power loss performance metric that better resembles end system objectives. Our results show that the proposed solution resulted in beam alignment power losses inferior to 3dB with respect the ideal alignment in all analyzed real-world scenarios. Also, for the considered scenarios, we verified 70\% overhead savings compared to exhaustive search with 95\% beam prediction accuracy guarantees. These results highlight the potential of position data to aid real-world beam alignment. 


\end{document}

%% file: input.tex
\usepackage{amsfonts}
\usepackage{times}
\usepackage{graphicx}
\usepackage{latexsym}
\usepackage{dsfont}
\usepackage{amssymb}
\usepackage{amsmath}
\usepackage{cite}
\usepackage{verbatim}
\usepackage{subfigure}

\newtheorem{algorithm}{Algorithm}

\def\bb0{{\mathbb{0}}}

\def\bb{{\mathbf{b}}}

\def\b0{{\mathbf{0}}}

\def\bbE{{\mathbb{E}}}

\def\sf0{{\mathsf{0}}}









%% file: Position-Aided_Beam_Prediction.bbl
\begin{thebibliography}{1}
	\providecommand{\url}[1]{#1}
	\csname url@samestyle\endcsname
	\providecommand{\newblock}{\relax}
	\providecommand{\bibinfo}[2]{#2}
	\providecommand{\BIBentrySTDinterwordspacing}{\spaceskip=0pt\relax}
	\providecommand{\BIBentryALTinterwordstretchfactor}{4}
	\providecommand{\BIBentryALTinterwordspacing}{\spaceskip=\fontdimen2\font plus
		\BIBentryALTinterwordstretchfactor\fontdimen3\font minus
		\fontdimen4\font\relax}
	\providecommand{\BIBforeignlanguage}[2]{{%
			\expandafter\ifx\csname l@#1\endcsname\relax
			\typeout{** WARNING: IEEEtran.bst: No hyphenation pattern has been}%
			\typeout{** loaded for the language `#1'. Using the pattern for}%
			\typeout{** the default language instead.}%
			\else
			\language=\csname l@#1\endcsname
			\fi
			#2}}
	\providecommand{\BIBdecl}{\relax}
	\BIBdecl
	
	\bibitem{itWillWork}
	T.~S. Rappaport, S.~Sun, R.~Mayzus, H.~Zhao, Y.~Azar, K.~Wang, G.~N. Wong,
	J.~K. Schulz, M.~Samimi, and F.~Gutierrez, ``Millimeter wave mobile
	communications for 5g cellular: It will work!'' \emph{IEEE Access}, vol.~1,
	pp. 335--349, 2013.
	
	\bibitem{geolocation-side-info}
	M.~Arvinte, M.~Tavares, and D.~Samardzija, ``Beam management in 5g nr using
	geolocation side information,'' in \emph{2019 53rd Annual Conference on
		Information Sciences and Systems (CISS)}, 2019, pp. 1--6.
	
	\bibitem{Elizabeth}
	S.~Rezaie, C.~N. Manchón, and E.~de~Carvalho, ``Location- and
	orientation-aided millimeter wave beam selection using deep learning,'' in
	\emph{IEEE International Conference on Communications}, 2020, pp. 1--6.
	
	\bibitem{9425522}
	Y.~Heng and J.~G. Andrews, ``Machine learning-assisted beam alignment for
	mmwave systems,'' \emph{IEEE Transactions on Cognitive Communications and
		Networking}, pp. 1--1, 2021.
	
	\bibitem{inverseFinger}
	V.~Va, J.~Choi, T.~Shimizu, G.~Bansal, and R.~W. Heath, ``Inverse multipath
	fingerprinting for millimeter wave v2i beam alignment,'' \emph{IEEE
		Transactions on Vehicular Technology}, vol.~67, pp. 4042--4058, 2018.
	
	\bibitem{robertInvited}
	Y.~Wang, A.~Klautau, M.~Ribero, M.~Narasimha, and R.~W. Heath, ``Mmwave
	vehicular beam training with situational awareness by machine learning,'' in
	\emph{2018 IEEE Globecom Workshops}, 2018, pp. 1--6.
	
	\bibitem{gaussian-noise-is-bad}
	A.~W. Soundy, B.~J. Panckhurst, and T.~C. Molteno, ``Enhanced noise models for
	gps positioning,'' in \emph{2015 6th International Conference on Automation,
		Robotics and Applications (ICARA)}, 2015, pp. 28--33.
	
	\bibitem{GitLink}
	M.~Joao, ``Position-aided beam prediction,''
	\url{github.com/jmoraispk/Position-Beam-Prediction}, 2021.
	
	\bibitem{DeepSense}
	\BIBentryALTinterwordspacing
	A.~Alkhateeb, G.~Charan, M.~Alrabeiah, T.~Osman, A.~Hredzak, N.~Srinivas, and
	M.~Seth, ``{DeepSense 6G}: Real-world multi-modal sensing and {CSI} dataset
	for {6G} deep learning research,'' \emph{available on arXiv}, 2021. [Online].
	Available: \url{https://www.DeepSense6G.net}
	\BIBentrySTDinterwordspacing
	
\end{thebibliography}
